\documentclass[12pt,showpacs,amsmath,amssymb,aps,prb]{revtex4-1}
\usepackage{graphicx}
\usepackage{subfigure}

\usepackage{dcolumn}
\usepackage{bm}

\begin{document}

\title{Casimir Forces and Graphene Sheets}

\author{D. Drosdoff and Lilia M. Woods}
\affiliation{Department of Physics, University of South Florida, Tampa FL 33620}

\begin{abstract}
The Casimir force between two infinitely thin parallel sheets in a setting of $N$ such sheets is found. 
The finite two-dimensional conductivities, which describe the dispersive and absorptive properties of each sheet, are taken into account, whereupon the theory is applied to interacting graphenes.  
By exploring similarities with in-plane optical spectra for graphite, the conductivity of graphene is modeled as a combination of Lorentz type oscillators. We find that 
the graphene transparency and the existence of a universal constant conductivity $e^2/(4\hbar)$ result in graphene/graphene Casimir interaction at large separations 
to have the same distance dependence as the one for perfect conductors but with much smaller magnitude.  

\end{abstract}

\maketitle

\section{Introduction \label{INTRO}}
The Casimir force is a fundamental quantum mechanical relativistic phenomenon which originates from the vacuum fluctuations of the electromagnetic field. It couples 
electrically neutral objects with or without permanent electric and/or magnetic moments.  In the case of two perfectly conducting infinite plates, the Casimir 
force depends only on the distance and two fundamental constants: the Plank's constant and the speed of light\cite{Casimir:1948,Mostepanenko:2009}. 

The Casimir force is 
of particular interest at the nanoscale. It has been shown that such a force is responsible for limiting the operation of many nanostructured 
devices, such as nano-electrical-mechanical and micro-electrical-mechanical systems by causing stiction, friction, or adhesion\cite{Chan1:2001, Chan:2001}. This has 
motivated devising experiments to detect the effect in various structures. Sophisticated techniques using torsional pendulum or atomic force microscope  have been used 
to measure the Casimir force between metallic and dielectric surfaces with high accuracy\cite{Lamoreaux:1997, Mohideen:1998}.The stability of many nanostructured materials, 
related devices and experimental settings has also been connected to dispersion forces originating from the Casimir effect. Graphitic nanostructures, such as 
graphenes (a single layer of graphite is graphene), carbon nanotubes (cylindrically rolled concentric graphenes), and graphene nanoribbons (finite width graphenes) are particular 
examples\cite{Dresselhaus_book,Neto:2009}.

Recently, single layers of graphite have been isolated using 
micromechanical cleavage\cite{Novoselov:2004,Novoselov:2005}. At present, graphene is one of the most interesting and most studied materials and has paved the way for future 
carbon based electronics. Many applications of graphene rely on its ability for continuously tuning its charge carrier density and mobility~\cite{Geim:2007}.  
This has allowed the development of new transistors operating at high frequencies.\cite{Lin:2010}  
Other applications are also very promising. Nanomechanical resonators, for example, are especially attractive due to their mechanical 
stability and high resonant frequency~\cite{Chen:2009}. 

Isolated graphenes also raise 
the possibility of studying the Casimir force between essentially two dimensional structures with peculiar dielectric response properties and uncovering further insights in 
the nature of the Casimir interaction. In addition, graphene/graphene or graphene/substrate mutual interactions are important components of many experimental settings. The 
Casimir force can be calculated using the Lifshitz theory, which takes into account the macroscopic dielectric response of the objects\cite{Klimchitskaya:2009}. This theory was also adapted for the 
case of two graphene sheets using an idealistic description of the dielectric permittivity by assuming a Drude type model\cite{Bordag:2006}. Such an approximation, however, 
does not take into account the electronic properties specific to graphene. Researchers have also considered the Casimir interaction between a graphene and a perfect 
conductor within a quantum field theory approach by 
using a more realistic representation of the graphene dielectric response\cite{Bordag:2009}, where low momentum electrons were modeled via the Dirac model.

The goal of this work is to present a theory for $N$ parallel infinitely thin sheets at different separations and take into account the specific 
optical properties characterized by the conductivity of each separate sheet. 
The objective is to apply the results to the case of parallel graphenes, in order to understand how the distance separation and peculiar characteristics of their 
dispersion properties manifest in their mutual Casimir force. Of particular interest is the relation of the force to the universal graphene conductivity, which 
provides a different representation of the effect in this peculiar system. Our method utilizes a quantum electrodynamical approach based on linear response theory. 
An essential part is the explicit calculation of the dyadic Green's function for this system and the inclusion of the finite conductivity of the separate sheets.  

The rest of the paper is organized as follow. In section \ref{EFF}, the electromagnetic field induced fluctuation stresses between two infinitely thin parallel plates is developed via linear response theory.  In section \ref{GF}, the dyadic Green's function for two sheets with frequency dependent conductivities is found.  In section \ref{FPS}, the Casimir force between two parallel sheets within N sheets is characterized via recursion relationships for the reflection coefficients.  Finally in section \ref{GL}, the theory is applied to find the Casimir force between graphenes.  The conclusions are described in section \ref{Conclusion}.

\section{Fluctuation forces\label{EFF}}

In classical electrodynamics, the force is calculated using the Maxwell stress tensor for the electromagnetic pressure 
 \begin{equation}
\overleftrightarrow{T}=\frac{1}{4\pi}\left({\bf E}{\bf E}-\frac{1}{2}E^2\overleftrightarrow{1}+
{\bf B}{\bf B}-\frac{1}{2}B^2\overleftrightarrow{1}\right),
\label{EFF1}
\end{equation}
where ${\bf E}$ and ${\bf B}$ are the electric and magnetic fields, respectively, and $\overleftrightarrow{1}$ is the unit matrix. In order to present a formalism to find the 
Casimir stresses, we consider the radiation-matter interaction Hamiltonian within the dipole approximation described by 
\begin{equation}
\delta {H}=-\int d^3r {\bf P}({\bf r},t)\cdot {\bf E}({\bf r},t),
\label{EFF2}
\end{equation}
where ${\bf P}$ is the polarization.

The quantum mechanical description is performed by simply substituting the classical fields with a symmetrized product\cite{Landau:1980} of quantum mechanical operators 
(denoted by hats in what follows). Then, using linear response theory\cite{Kubo:1957}, the perturbed electric  fields due to an external polarization source field become
\begin{eqnarray}
\delta{\bf E}({\bf r},t)&=&\int_{-\infty}^tdt'\frac{i}{\hbar}\int d^3r\overleftrightarrow{G}({\bf r},{\bf r'},t-t')\cdot{\bf  P}({\bf r'},t'), \\
\overleftrightarrow{G}({\bf r},{\bf r'},t-t')&=&\frac{i}{\hbar}\langle[{\bf \hat{E}}({\bf r},t),{\bf \hat{E}}({\bf r'},t')]\rangle,
\label{EFF3}
\end{eqnarray}
where $\langle \hat{O}\rangle=Tr(\hat{\rho}\hat{O})$, $\hat{\rho}=exp([F_0-\hat{H}_0]/k_BT)$ is the statistical matrix operator in the canonical ensemble,
$F_0$ is the free energy, $\hat{H}_0$ is the unperturbed Hamiltonian, $\overleftrightarrow{G}$ is the Dyadic Green's function, $k_B$ is Boltzmann's constant, and $T$ is the 
temperature.  Note that $\overleftrightarrow{G}({\bf r}, {\bf r}',t,t')=\overleftrightarrow{G}({\bf r}-{\bf r}',t-t')$ from the homogeneity of space-time.

Define the structure functions $\overleftrightarrow{S}_{EE}$ and $\overleftrightarrow{S}_{BB}$ using Fourier transforms as follows
\begin{eqnarray}
\frac{1}{2}\langle[{\bf \hat{E}}({\bf r}, t){\bf \hat{E}}({\bf r'},t')+{\bf \hat{E}}({\bf r'}, t'){\bf \hat{E}}({\bf r},t)\rangle
&=&\int \frac{d\omega}{2\pi} \overleftrightarrow{S}_{EE}({\bf r},{\bf r'},\omega)e^{-i\omega(t-t')}, \nonumber \\
\frac{1}{2}\langle[{\bf \hat{B}}({\bf r}, t){\bf \hat{B}}({\bf r'},t')+{\bf \hat{B}}({\bf r'}, t'){\bf \hat{B}}({\bf r},t)\rangle
&=&\int \frac{d\omega}{2\pi} \overleftrightarrow{S}_{BB}({\bf r},{\bf r'},\omega)e^{-i\omega(t-t')}.
\label{EFF5}
\end{eqnarray}
Then using the fluctuation dissipation theorem\cite{Knoll:2001,Tomas:2002}, one can express the structure function $\overleftrightarrow{S}_{EE}$ in terms of the 
imaginary part of the frequency dependent Green's function at finite temperature as
\begin{equation}
\overleftrightarrow{S}_{EE}({\bf r},{\bf r'},\omega) = \hbar\times\Im m\overleftrightarrow{G}({\bf r},{\bf r'},\omega)\coth\left(\frac{\hbar\omega}{2k_BT}\right).
\label{EFF6}
\end{equation}
Similarly, the correlations in the magnetic field may be found from Faraday's law by taking the curl of the electric field. These are   
expressed in terms of the function $\overleftrightarrow{S}_{BB}$ by
\begin{equation}
\overleftrightarrow{S}_{BB}({\bf r},{\bf r'},\omega)=\hbar\frac{c^2}{\omega^2}\Im m\nabla\times\overleftrightarrow{G}({\bf r},{\bf r'},\omega)
\times\nabla'\coth\left(\frac{\hbar\omega}{2k_BT}\right),
\label{EFF7}
\end{equation}
where $\times\nabla'$ is the curl taken on the Green's function with respect to the prime coordinates.
Therefore, from Eqs.(\ref{EFF1},\ref{EFF5},\ref{EFF6},\ref{EFF7}) the temperature dependent quantum electromagnetic stress is found via the Green's function\cite{Raabe:2003},
\begin{eqnarray}
&&\overleftrightarrow{T}=\frac{1}{4\pi}\left(\overleftrightarrow{T}_1+\overleftrightarrow{T}_2
-\frac{\overleftrightarrow{1}}{2}Tr\left[\overleftrightarrow{T}_1+\overleftrightarrow{T}_2\right]
\right),\nonumber \nonumber\\
&& \overleftrightarrow{T}_1=\lim_{{\bf r}\rightarrow {\bf r'} }\int\overleftrightarrow{S}_{EE}({\bf r},{\bf r'},\omega)\frac{d\omega}{2\pi},\nonumber\\ 
&& \overleftrightarrow{T}_2=\lim_{{\bf r}\rightarrow {\bf r'}}\int\overleftrightarrow{S}_{BB}({\bf r},{\bf r'},\omega)\frac{d\omega}{2\pi}.
\label{EFF8}
\end{eqnarray}
Finally, the identity $\coth(x)=\sum_{n=-\infty}^{\infty}\frac{x}{x^2+n^2\pi^2}$ allows one to obtain
\begin{eqnarray}
&& \overleftrightarrow{T}_1=\lim_{{\bf r}\rightarrow {\bf r'}}k_BT\sum_{n=-\infty}^\infty\overleftrightarrow{G}({\bf r},{\bf r'},i\omega_n),\nonumber\\ 
&& \overleftrightarrow{T}_2=\lim_{{\bf r}\rightarrow {\bf r'}}k_BT\sum_{n=-\infty}^\infty\frac{c^2}{(i\omega_n)^2}\nabla\times
\overleftrightarrow{G}({\bf r},{\bf r'},i\omega_n)\times\nabla',
\label{EFF10}
\end{eqnarray}
where $\omega_n=2\pi n k_BT/\hbar$ are the Matsubara frequencies.

\section{Green's Function for three layers\label{GF}}

In order to find the Casimir force from Eq.(\ref{EFF8}) for a specific structure, one needs to 
calculate the Green's function. The Green's function obeys the following equation
\begin{equation}
\left[\nabla\times\nabla -\frac{\omega^2}{c^2}\right]\overleftrightarrow{G}({\bf r},{\bf r'},\omega)=4\pi\frac{\omega^2}{c^2}\delta({\bf r}-{\bf r'}),
\label{GF1}
\end{equation}
which is found from the solution of Maxwell's equations\cite{Hanson:2008}.

The system under consideration consists of parallel, infinitely thin sheets located in a vacuum.
Each sheet is positioned at the boundary between two adjacent layers, and it is specified by a two-dimensional, isotropic conductivity  
\begin{equation}
\overleftrightarrow{\sigma}_{j,j+1}=
\left(
\begin{array}{ccc}
\sigma_{j,j+1} & 0 & 0 \\
0 & \sigma_{j,j+1} & 0 \\
0 & 0 & 0 \end{array}
\right)
\label{GF6.1}
\end{equation}
where $j,j+1$ is the boundary between two adjacent layers.  $\sigma_{j,j+1}$ accounts for the specific finite absorptive and dispersive optical properties of each sheet. 

\begin{figure}[ht]
\begin{centering}
\includegraphics[scale=0.35]{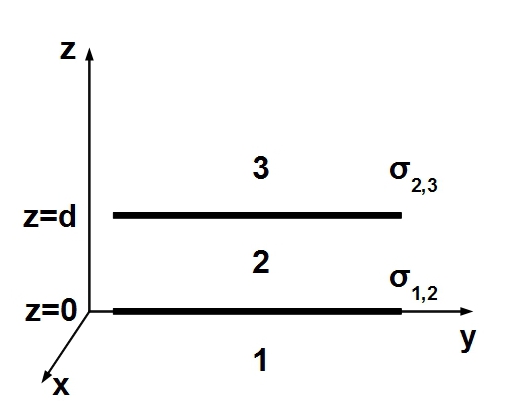}
\caption{The free space regions denoted as $1,2,3$ separated by two infinitely thin sheets extending in the $x-y$ plane and separated by a distance $d$ in the $z$-direction.  
The infinitely thin sheets are located at $z=0$ and $z=d$.  The conductivities are also denoted. }
\label{sheets}
\end{centering}
\end{figure}
In general, the Green's function may be split in two terms

\begin{equation}
\overleftrightarrow{G}^{lj}=\overleftrightarrow{G}^{j}_{0}+\overleftrightarrow{G}_s^{lj}
\end{equation}
where $\overleftrightarrow{G}^{lj}$ is the Green's function in region $l$ with a source in region $j$, $\overleftrightarrow{G}^{j}_{0}$ is the free Green's function from a 
point-like source placed in layer $j$ without any boundaries, and $\overleftrightarrow{G}_{s}^{lj}$  is the scattering Green's function in region $l$ with a source in region $j$. 
The free Green's function is later dropped from the calculation of the stress tensor since the Casimir effect does not exist in homogeneous space.

The calculation of the force between planar sheets given in this work is based on a procedure using the generalized Fresnel reflection coefficients. The method involves 
the explicit form of the dyadic Green's function for a system of two parallel planar sheets\cite{Knoll:2001,Tomas:2002}. Thus we first calculate 
$\overleftrightarrow{G}({\bf r},{\bf r'},\omega)$ for the three layer/two sheet structure shown in Fig.\ref{sheets}. 

Since the system has a planar geometry, we use the general dyadic form of $\overleftrightarrow{G}$ in terms of the following orthogonal functions\cite{Tai:1993, Cheng:1986}
\begin{equation}
{\bf M}({\bf k})=\nabla\times\left[\hat{{\bf z}}\phi\right]\ \ \ 
{\rm and}\ \ \  {\bf N}({\bf k})=\frac{1}{k}\nabla\times{\bf M}({\bf k}), 
\label{GF2}
\end{equation}
where $\phi=exp(i{\bf k}\cdot {\bf r})$ with ${\bf k}$ being the wave vector of the electromagnetic excitations. These obey the orthogonality relations
\begin{eqnarray}
&&\int dV {\bf M}({\bf k})\cdot{\bf N}(-{\bf k'})=0\nonumber \\
&&\int dV {\bf M}({\bf k})\cdot{\bf M}(-{\bf k'})=(2\pi)^3 k_{\perp}^{2} \delta({\bf k}-{\bf k'})\nonumber \\
&&\int dV {\bf N}({\bf k})\cdot{\bf N}(-{\bf k'})=(2\pi)^3 k_{\perp}^{2} \delta({\bf k}-{\bf k'})
\label{GF3}
\end{eqnarray}
where $k_{\perp}^{2}=k_x^2+k_y^2$ and the integration is carried over space. 
The bulk Green's function is found to be of the following form:
\begin{eqnarray}
&&\overleftrightarrow{G}^j_0({\bf r},{\bf r'},\omega)=-4\pi\delta({\bf r}-{\bf r'}){\bf z}{\bf z}+\frac{i\omega^2}{c^2}\int\frac{dk^2_\perp}{2\pi k^2_\perp h}
\left[{\bf M}(h){\bf M'}(-h) +{\bf N}(h){\bf N'}(-h)\right],\nonumber\\
&&(z-z'>0)\\
&&\overleftrightarrow{G}^j_0({\bf r},{\bf r'},\omega)=-4\pi\delta({\bf r}-{\bf r'}){\bf z}{\bf z}+\frac{i\omega^2}{c^2}\int\frac{dk^2_\perp}{2\pi k^2_\perp h}
\left[{\bf M}(-h){\bf M'}(h) +{\bf N}(-h){\bf N'}(h)\right],\nonumber \\
&& (z-z'<0),
\label{GF6}
\end{eqnarray}
\begin{eqnarray}
&&{\bf M}(h)=i(k_y{\bf\hat{x}}-k_x{\bf\hat{y}})e^{i({\bf k}_\perp\cdot{\bf r}_\perp+hz)},\ \ \ {\bf M}(-h)=i(k_y{\bf\hat{x}}-k_x{\bf\hat{y}})e^{i({\bf k}_{\perp}\cdot {\bf r}_{\perp}-hz)}\nonumber \\
&&{\bf M'}(-h)=-i(k_y{\bf\hat{x}}-k_x{\bf\hat{y}})e^{-i({\bf k}_\perp\cdot {\bf r'}_\perp+hz')},\ \ \ {\bf M'}(h)=-i(k_y{\bf\hat{x}}-k_x{\bf\hat{y}})e^{-i({\bf k}_\perp\cdot {\bf r'}_\perp-hz')},
\label{GF5}
\end{eqnarray}
where $h=\sqrt{\omega^2/c^2-k_\perp^2}$ and ${\bf r}_\perp=x {\bf\hat{x}}+y {\bf\hat{y}}$.
Similarly, ${\bf N}(\pm h)$,  ${\bf N'}(\pm h)$ are defined via Eq.(\ref{GF2}). 

In this way, the bulk Green's function is expressed in terms of a linear combination of transverse electric ({\bf M}-term) and transverse magnetic ({\bf N}-term) modes. The 
scattering part of the Green's function
is also sought in terms of the orthogonal functions {\bf M} and {\bf N}, and it is found from the boundary conditions for the tangential electric and magnetic fields across each 
plane. 

The continuity of the electric field across each boundary from Fig(\ref{sheets}) is expressed as
\begin{eqnarray}
\hat{{\bf z}}\times\left[\overleftrightarrow{G}^{22}({\bf r},{\bf r'},\omega)-\overleftrightarrow{G}^{12}({\bf r},{\bf r'},\omega)\right]=0,\ \ \ (z=0),\nonumber\\
\hat{{\bf z}}\times\left[\overleftrightarrow{G}^{32}({\bf r},{\bf r'},\omega)-\overleftrightarrow{G}^{22}({\bf r},{\bf r'},\omega)\right]=0,\ \ \ (z=d),
\label{GF7}
\end{eqnarray}
while the tangential components of the magnetic field give rise to surface currents, which may be written as
\begin{eqnarray}
&&\hat{{\bf z}}\times\left[\nabla\times\overleftrightarrow{G}^{22}({\bf r},{\bf r'},\omega)
-\nabla\times\overleftrightarrow{G}^{12}({\bf r},{\bf r'},\omega)\right]=
\frac{4\pi i\omega}{c^2}\overleftrightarrow{\sigma}_{1,2}\cdot\overleftrightarrow{G}^{12}({\bf r},{\bf r'},\omega),\ \ \ (z=0)\nonumber\\
&&\hat{{\bf z}}\times\left[\nabla\times\overleftrightarrow{G}^{32}({\bf r},{\bf r'},\omega)
-\nabla\times\overleftrightarrow{G}^{22}({\bf r},{\bf r'},\omega)\right]=
\frac{4\pi i\omega}{c^2}\overleftrightarrow{\sigma}_{2,3}\cdot\overleftrightarrow{G}^{32}({\bf r},{\bf r'},\omega),\ \ \  (z=d). 
\label{GF8}
\end{eqnarray}
The surface conductivities account for the finite absorption properties of the infinitely thin sheets in the material. The solution of the dyadic Green's function in 
the different regions may be written in terms of the reflection coefficients for the transverse electric(TE) waves and the transverse magnetic(TM) waves.  For the system 
considered in Fig.(\ref{sheets}), these are
\begin{eqnarray}
&&\rho^+_E=-\frac{2\pi\omega\sigma_{2,3}/(hc^2)}{1+2\pi\omega\sigma_{2,3}/(hc^2)},
\ \ \ \rho^-_E=-\frac{2\pi\omega\sigma_{1,2}/(hc^2)}{1+2\pi\omega\sigma_{1,2}/(hc^2)}
\nonumber \\
&&\rho^+_B=\frac{2\pi\sigma_{2,3} h/\omega}{1+2\pi\sigma_{2,3} h/\omega},\ \ \ 
\rho^-_B=\frac{2\pi\sigma_{1,2} h/\omega}{1+2\pi\sigma_{1,2} h/\omega}
\label{GF8.1}
\end{eqnarray}
where $(+)$ superscript defines the top plate and $(-)$ superscript defines the bottom plate. 

Making the definition $\Omega_{E,B}=1-\rho_{E,B}^+\rho_{E,B}^- e^{2ihd}$,
the scattering Green's function in region two is 
\begin{eqnarray}
&&\overleftrightarrow{G}^{22}_s({\bf r},{\bf r'})=\frac{i\omega^2}{c^2}\int\frac{dk^2_\perp}{2\pi k^2_\perp h}[\nonumber \\
&&\frac{\rho_E^-}{\Omega_E}{\bf M}(h){\bf M'}(h)+ \frac{\rho_E^+\rho_E^-e^{2ihd}}{\Omega_E}{\bf M}(h){\bf M'}(-h)\nonumber \\
&&+\frac{\rho_E^+\rho_E^-e^{2ihd}}{\Omega_E}{\bf M}(-h){\bf M'}(h) +\frac{\rho_E^+e^{2ihd}}{\Omega_E}{\bf M}(-h){\bf M'}(-h) \nonumber \\
&&+\frac{\rho_B^-}{\Omega_B}{\bf N}(h){\bf N'}(h)+\frac{\rho_B^+\rho_B^-e^{2ihd}}{\Omega_B}{\bf N}(h){\bf N'}(-h)\nonumber \\
&&+\frac{\rho_B^+\rho_B^-e^{2ihd}}{\Omega_B}{\bf N}(-h){\bf N'}(h)+\frac{\rho_B^+e^{2ihd}}{\Omega_B}{\bf N}(-h){\bf N'}(-h)].
\label{GF9}
\end{eqnarray} 
Similarly, the scattering Green's function in region one is
\begin{eqnarray}
&&\overleftrightarrow{G}^{11}_s({\bf r},{\bf r'})=\frac{i\omega^2}{c^2}\int\frac{dk^2_\perp}{2\pi k^2_\perp h}[\nonumber \\
&&\left(\frac{\rho_E^-+(\rho_E^++2\rho_E^-\rho_E^+)e^{2ihd}}{\Omega_E}\right){\bf M}(-h){\bf M'}(-h)\nonumber \\
&&+\left(\frac{\rho_B^-+(\rho_B^+-2\rho_B^-\rho_B^+)e^{2ihd}}{\Omega_B}\right){\bf N}(-h){\bf N'}(-h)].
\label{GF10}
\end{eqnarray}
The other scattering Green's functions are found from the boundary conditions specified in Eqs.(\ref{GF7},\ref{GF8}).

\section{Force between N parallel sheets\label{FPS}}

The Casimir force per unit area exerted on each planar sheet Fig.(\ref{sheets}) is obtained by evaluating the zz-component of the difference in the stress between the regions 
above and below that sheet. For example, the force on the bottom one is calculated by taking
\begin{equation}
T_b=\left[T_{zz}^{22}(z)-T_{zz}^{11}(z)\right]_{z=0},
\label{FPS1}
\end{equation} 
where $T_b$ is the force per unit area on the bottom plate, $T_{zz}^{22}$ is the zz-component of the stress in 
region $2$ given a fluctuating source in region $2$, and $T_{zz}^{11}$  is the zz-component of the stress in 
region $1$ given a fluctuating source in region $1$.
The force per unit area on the top plate is equal and opposite to that on the bottom one, ie, $T_t=-T_b$.  Combining the results in previous sections with 
Eq.(\ref{FPS1}) one obtains 
\begin{equation}
T_b=-\frac{ik_BT}{2\pi}\sum_{n=-\infty}^{\infty}\int_0^\infty h(i\omega_n) k_\perp dk_{\perp}
\left\{\left[\frac{e^{-2ih(i\omega_n)d}}{\rho_E^+(i\omega_n)\rho_E^-(i\omega_n)}-1\right]^{-1}+
\left[\frac{e^{-2ih(i\omega_n)d}}{\rho_B^+(i\omega_n)\rho_B^-(i\omega_n)}-1\right]^{-1}\right\},
\label{FPS2}
\end{equation}
where $h(i\omega_n)=i\sqrt{(\omega_n/c)^2+k_\perp^2}$. The expression for $T_{b}$ can be used to calculate the temperature dependent interaction between any two infinitely 
thin plates in vacuum providing that the explicit conductivities are known. One notes that the largest contribution\cite{Abrikosov:1963} of Eq.(\ref{FPS2}) comes 
from $\omega_n d/c\approx 1$. Therefore, when $\hbar c/(2\pi k_BTd)\gg 1$, the sum above is determined by the large $n$ value terms and it 
can be transformed into an integral with differential $d\omega_n=2\pi k_BT(dn)/\hbar$. 
For $T=0$, such a representation is exact, but for $T\leq 300\ K$, for example, $d$ should be less than a micrometer. At distances comparable 
to the thermal quantum coherence wavelength $\lambda_T=\hbar c/(k_BT)\approx 7\ \mu m$, classical thermal fluctuations become important and $\hbar c/(2\pi k_BTd)\gg 1$ is not 
valid any more. In this case, the $n=0$ and small $n$ terms in the sum become important, and the $d\omega_n \sim dn$ transformation cannot be justified. 
Here, we will assume that we are in a regime where the integral representation is valid. 

\begin{figure}[ht]
\begin{centering}
\includegraphics[scale=0.35]{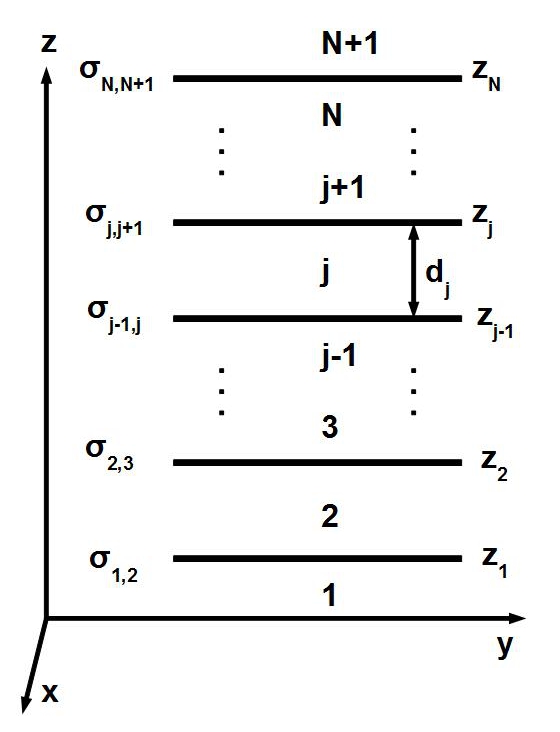}
\caption{$N$ infinitely thin sheets located in free space and separated by distances $d_j$. The sheets extend in the $x-y$ plane. Their conductivities and positions along the 
$z$-axis are also shown. }
\label{nsheets}
\end{centering}
\end{figure}

The result for the system of three layer/two sheet system can be used to obtain the force per unit area in the case of a stack of $N$ parallel sheets. Consider the $j$-th 
layer in Fig.(2). One  
realizes that the Casimir force results from the infinite optical reflecting and transmitting paths due to the scattering from all the sheets above and below layer $j$. 
This is described by the effective reflection from below $\rho_{j,E,B}^{-}$ and from above $\rho_{j,E,B}^{+}$ in layer $j$. Then combining Eq.(\ref{FPS2}) with the 
condition $\hbar c/(2\pi k_BTd)\gg 1$ and the integral representation that follows, the stress on the bottom sheet may be written as 
\begin{equation}
T_{bj}=-\frac{i\hbar}{2\pi^2}\int_0^\infty k_\perp dk_\perp\int_0^\infty d\omega h(i\omega) \left\{\left[\frac{e^{-2ih(i\omega)d_j}}{\rho_{Ej}^+(i\omega)
\rho_{Ej}^-(i\omega)}-1\right]^{-1}+
\left[\frac{e^{-2ih(i\omega)d_j}}{\rho^{+}_{Bj}(i\omega)\rho^{-}_{Bj}(i\omega)}-1\right]^{-1}\right\}.
\label{SPS4}
\end{equation}
The temperature dependence does not appear explicitly in the force any more. It is accounted for indirectly through the temperature dependent optical properties of the sheets.  
$\rho_{j,E,B}^{\pm}$ can be found via an iterative procedure using a simple recursion relation. Consider the three 
layers denoted as $(j-1),j,(j+1)$ by themselves. Due to the infinite optical paths they can be expressed as\cite{Tomas:2002,Ellingsen:2007} 
\begin{eqnarray}
\rho_{E,j-1,j,j+1}&=&\rho_{E,j-1,j}+t_{E,j-1,j}\rho_{E,j,j+1}t_{E,j,j-1}e^{2ihd_j}+\nonumber \\
&&t_{E,j-1j}\rho_{E,j,j+1}t_{E,j,j-1}e^{4ihd_j}\rho_{E,j,j-1}\rho_{E,j,j+1}+\ldots,\nonumber \\
\rho_{E,j-1,j,j+1}&=&\rho_{E,j-1,j}+ t_{E,j-1,j}\rho_{E,j,j+1}t_{E,j,j-1}e^{2ihd_j}\sum_{n=0}^\infty\left[\rho_{E,j,j-1}\rho_{E,j,j+1}e^{2ihd_j}\right]^n,
\label{SPS5}
\end{eqnarray}
where $\rho_{E,j-1,j}$ is a single sheet reflection coefficient from layer $j-1$ to layer $j$ and $t_{E,j-1,j}$ is the coefficient of transmission for a single sheet from layer 
$j-1$ to layer $j$.  Given that $t_{E,j-1,j}=t_{E,j,j-1}$, $\rho_{E,j-1,j}=\rho_{E,j,j-1}$ and 
$1+\rho_{E,j-1,j}=t_{E,j-1,j}$ one obtains
\begin{equation}
\rho_{E,j-1,j,j+1}=\frac{\rho_{E,j-1,j}+(\rho_{E,j,j+1}+2\rho_{E,j,j+1}\rho_{E,j,j-1})e^{2ihd_j}}{1-\rho_{E,j,j-1}\rho_{E,j,j+1}e^{2ihd_j}}.
\label{SPS6}
\end{equation}
Similarly, one finds the reflection coefficient for the TM modes given that $t_{B,j-1,j}=t_{B,j,j-1}$, $\rho_{B,j-1,j}=\rho_{B,j,j-1}$ and $1-\rho_{B,j-1,j}=t_{B,j-1,j}$,
\begin{equation}
\rho_{B,j-1,j,j+1}=\frac{\rho_{B,j-1,j}+(\rho_{B,j,j+1}-2\rho_{B,j,j+1}\rho_{B,j,j-1})e^{2ihd_j}}{1-\rho_{B,j,j-1}\rho_{B,j,j+1}e^{2ihd_j}}.
\label{SPS6.1}
\end{equation}
This provides a straight forward method for calculating the reflection coefficients in any vacuum layer in a stack of $N$ parallel sheets. Suppose $j-1,j,j+1$ are a part of the 
system shown in Fig.(\ref{nsheets}). Starting from layer 1, one finds the reflection coefficients between the first and second layers using Eq.(\ref{GF8.1}). 
Invoking Eqs.(\ref{SPS6},\ref{SPS6.1}) recursively by treating 
the first two sheets as one, the reflection coefficient in the third layer is found. This is repeated until layer $j$ is reached giving the effective reflection from below. A 
similar procedure is applied to find the reflection from all sheets from above layer $j$, but starting from the top $N+1$ layer in Fig.(\ref{nsheets}).  

As an example of the use of the recursion procedure, a four layered system will be considered, which corresponds to the bottom four layers of Fig.(\ref{nsheets}).
Using Eqs.(\ref{SPS6},\ref{SPS6.1}), the generalized reflection coefficients in layer 2 are expressed as
\begin{eqnarray}
&&\rho_{E2}^+=\rho_{E234}=\frac{\rho_{E23}+(\rho_{E34}+2\rho_{E23}\rho_{E34})e^{2ihd_3}}{1-\rho_{E23}\rho_{E34}e^{2ihd_3}},\nonumber \\
&&\rho_{B2}^+=\rho_{B234}=\frac{\rho_{B23}+(\rho_{B34}-2\rho_{B23}\rho_{B34})e^{2ihd_3}}{1-\rho_{B23}\rho_{B34}e^{2ihd_3}}, \nonumber\\
&&\rho_{E2}^-=\rho_{E21}, \ \ \ 
\rho_{B2}^-=\rho_{B21}
\label{SPS7}
\end{eqnarray}
Note that $\rho^{-}_{E,B,2}$ are actually  $\rho^{-}_{E,B}$ from Eq.(\ref{GF8.1}), since the system below layer 2 is the same as the one from Fig.(1). The reflection coefficients 
can be substituted in Eq.(\ref{SPS4}) yielding the Casimir force per unit area on the bottom plate in the four layer/three plate system.

\section{Casimir interaction between graphenes\label{GL}}

Before the graphene planes are considered, it is useful to calculate the interaction for the limit of infinitely conducting planar sheets. 
In that case, the conductivity becomes $\sigma\rightarrow\infty$ 
yielding $\rho_{Ej}^\pm\rightarrow-1$ and $\rho_{Bj}^\pm\rightarrow 1$. Thus using Eq.(\ref{SPS4}) we recover the well known result\cite{Casimir:1948} for the magnitude of the attraction 
between two parallel perfectly conducting plates separated by a distance $d$  
\begin{equation}
|T_0|=\frac{\hbar c\pi^2}{240 d^4}.
\label{SPS10}
\end{equation}

\subsection {Universal conductivity}

Further, we apply the results for infinitely thin sheets obtained in section \ref{FPS} to calculate the Casimir force between two graphenes. Researchers in the past have 
considered graphene as an infinitely thin sheet, and have shown that this is a reasonable approximation for distances greater than a few times the interlayer graphite 
separation~\cite{Bordag:2006,Bordag:2009,Santos:2009}. The emphasis now is to specify the graphene conductivity.  
It has been predicted\cite{Gusynin:2006,Falkovsky:2007} and found experimentally\cite{Nair:2008}, that over a relatively wide range of photon energies 
(up to $3\ eV$), the graphene conductivity is approximately constant given by the value of $\sigma_0=e^2/(4\hbar)$. This peculiar effect is closely 
related to the energy band structure of graphene, as shown in the Appendix. 

Given a constant conductivity $\sigma_0$ for two parallel sheets separated by a distance $d$, Eq.(\ref{SPS4}) may be written in the following form  
\begin{eqnarray}
&&|T_g|=\frac{3\hbar c}{16\pi^2 d^4}\sum_{n=1}^\infty\left(\frac{1}{n^4}\right)\left[F(\sigma_0,n)+G(\sigma_0,n)\right],\nonumber \\
&&F(\sigma_0,n)=\frac{c}{2\pi\sigma_0}\beta(\frac{2\pi\sigma_0/c}{1+2\pi\sigma_0/c},2n+1,-1),\nonumber \\
&&G(\sigma_0,n)=\frac{2\pi\sigma_0}{c}\left(\frac{1}{2n-1}
\left[1-\left(\frac{2\pi\sigma_0/c}{1+2\pi\sigma_0/c}\right)^{(2n-1)}\right]\right)
\label{CC1}
\end{eqnarray}

where
\begin{equation}
\beta(x,a,b)=\int_0^xt^{a-1}(1-t)^{b-1}dt
\label{CC2} 
\end{equation}
is the incomplete beta function\cite{Abramowitz:1964}.

For graphene, however, $2\pi\sigma_0/c\ll 1$. Then Eq.(\ref{CC1}) is approximated by taking the first term in the sum, which reduces to
\begin{equation}
|T_g|\approx\frac{3\hbar c}{8\pi^2 d^4}\left[1-\frac{c}{2\pi\sigma_0}\ln(1+2\pi\sigma_0/c)\right].
\label{CC3}
\end{equation}
Further expansion of the ln-function and inserting $\sigma_0$ gives $|T_g|\approx\frac{3\hbar c}{16\pi^{2}d^{4}}\frac{2\pi\sigma_0}{c}=\frac{3e^{2}}{32\pi d^{2}}$. Thus the 
leading term in the force does not depend explicitly on the Plank's constant and speed of light any more. This is a remarkable result originating from the particular value 
of the graphene conductivity. We note that the approximate result from Eq. (\ref{CC3}) is fairly accurate 
since it differs by less than $2\%$ from the numerical integration of the exact result (Eq.(\ref{SPS4})).  

It is also interesting to see the similarities and differences between $T_g$ and $T_0$. In particular, the 
distance dependence of the Casimir force is the same as the one between two perfect conductors. However, comparing their 
values gives $T_g/T_0\approx0.00538$. Thus the graphene Casimir interaction is much smaller in magnitude than the interaction between perfect conductors. This is directly related to 
the transparency of the graphene system, reflected in its small constant conductivity value. 
The interaction between a perfectly conducting plate and graphene can also be calculated via Eq.(\ref{SPS4}).  In this case the force is much larger as compared to the one between two graphenes.  Given the constant $\sigma_0$ for graphene and $\sigma\rightarrow\infty$ for a perfect metal, one finds $T/T_0\approx 0.025$.

The general formula from Eq. (\ref{SPS4}) allows one to consider the Casimir interaction in a system with more than two parallel graphenes.  We can estimate the force between two 
graphenes in a three graphene setting. Using Eq. (\ref{SPS4}), the force is obtained as 
\begin{equation}
|T|=\frac{\hbar c}{2\pi^2}\frac{1}{16d_3^4}\int_1^\infty\frac{dp}{p^2}\int_0^\infty x^3dx\sum_{n=1}^\infty e^{-xn}
\left[(\rho_{E21}\rho_{E234})^n+(\rho_{B21}\rho_{B234})^n\right],
\label{GL2}
\end{equation}
where the reflection coefficients were defined in Eq.(\ref{SPS7}). Since $2\pi\sigma_0/c\ll 1$, the dominant contribution comes from the $n=1$ term:
\begin{equation}
|T|\approx\frac{3\hbar c}{16\pi^2d_3^4}\left\{\frac{2\pi\sigma_0}{c}-\frac{2}{3}\left(\frac{2\pi\sigma_0}{c}\right)^2\left[1+\left(\frac{d_2}{d_2+d_3}\right)^4\right]\right\}.
\label{GL4}
\end{equation}
Thus the interaction is determined mainly by the two adjacent graphenes in the three graphene system, and it is affected little (only to order $\sigma_0^2$) 
by the presence of the third one again due to their transparency. 

\subsection{Other models for the conductivity}

The low energy graphene band structure has been very successful in explaining experimentally observed properties at various temperatures\cite{Gusynin:2006, Falkovsky:2007}. Since the Casimir interaction 
at larger separations is determined by that regime (corresponding to optical excitations less than 3 eV),  
one concludes that its qualitative features cannot be an exception. Nevertheless, as the graphenes are brought closer, the 
presence of the higher energy bands besides the ones closest to the Fermi level also needs to be considered.

\begin{figure}[ht]
\centering
\subfigure{
\includegraphics[scale=0.35]{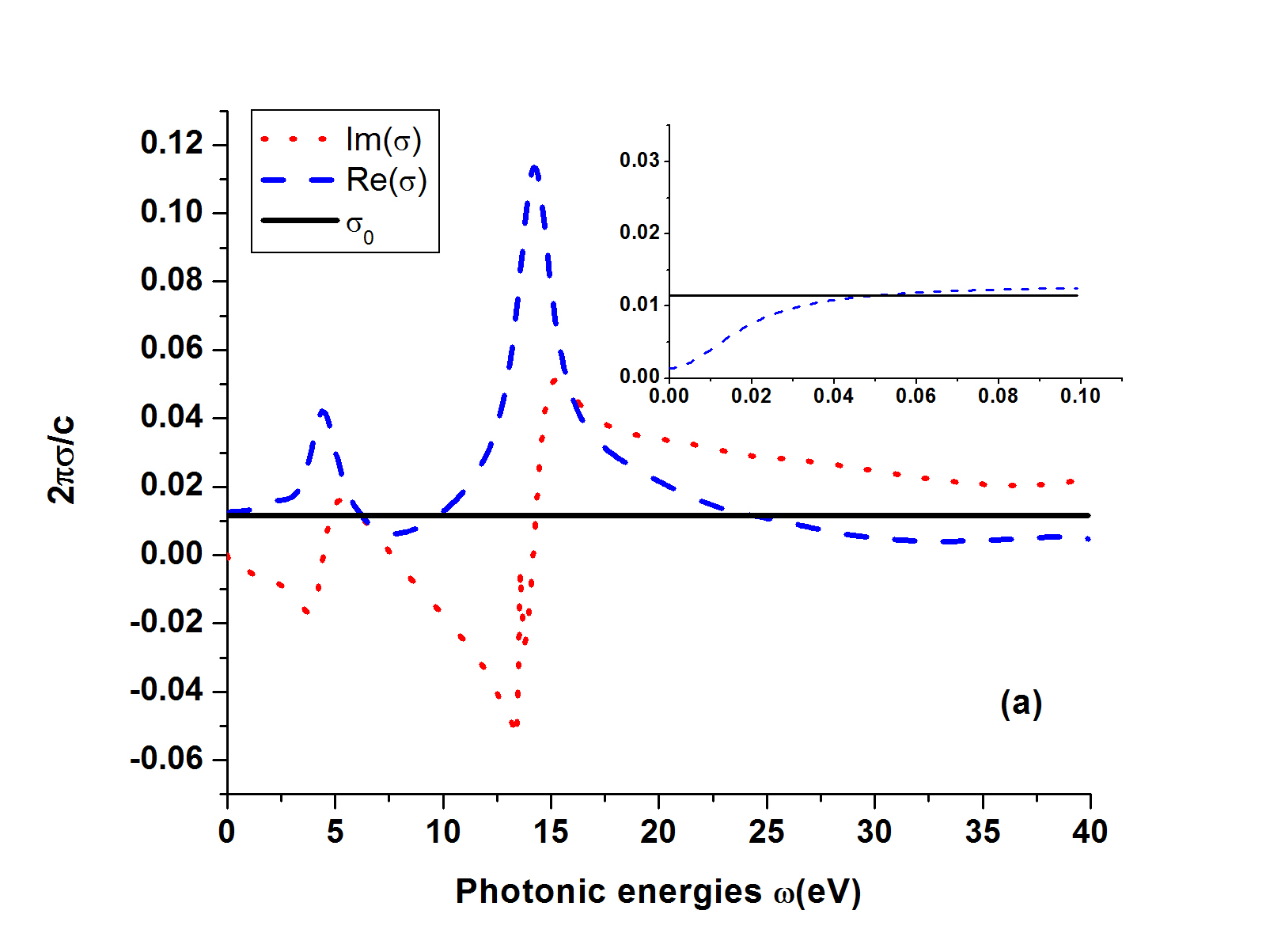}
\label{condgraph}
}
\subfigure{
\includegraphics[scale=0.35]{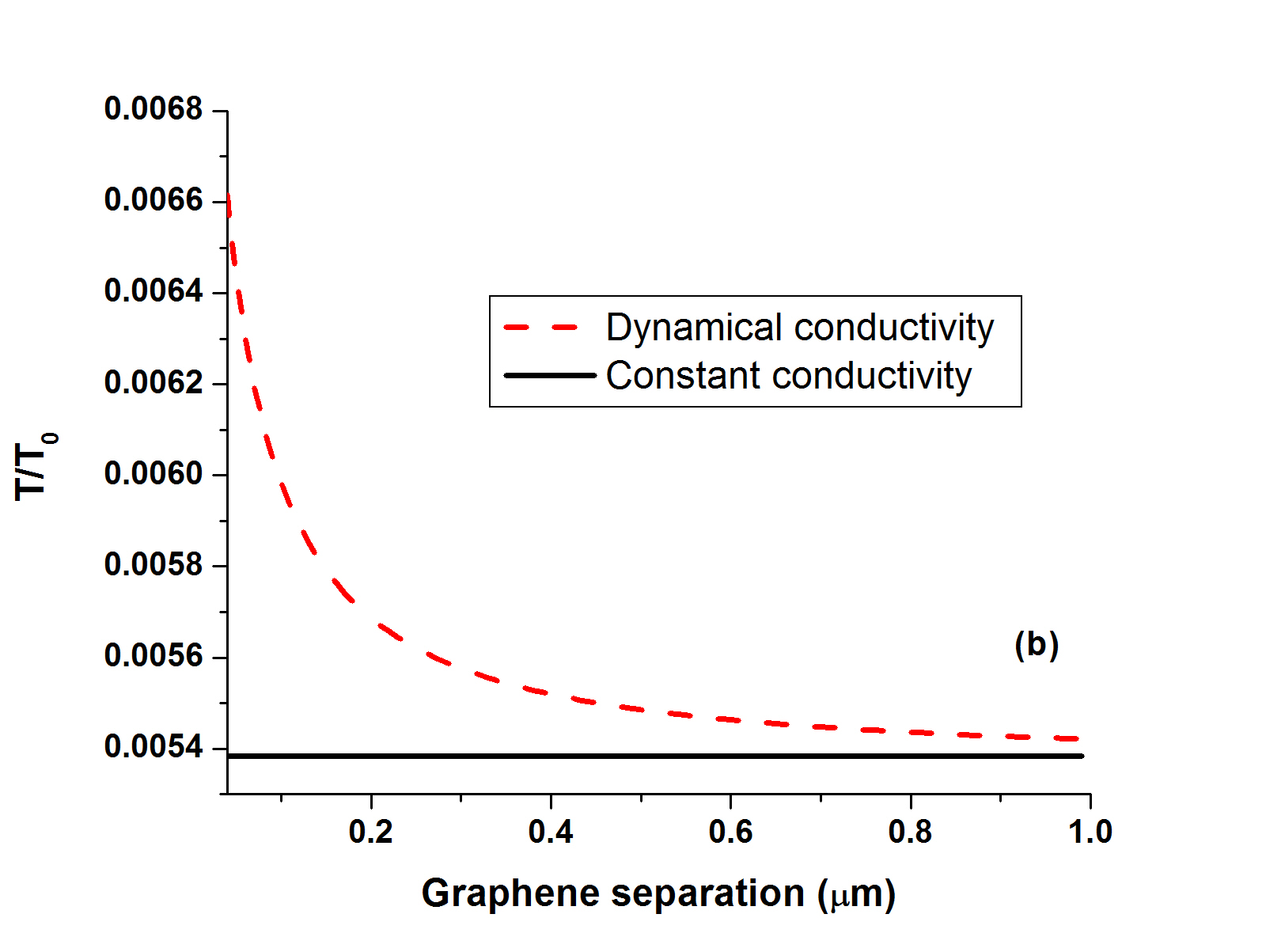}
\label{force2plates}
}
\label{figures}
\caption{(Color online)  a) $Re$ and $Im$ parts of the dynamical conductivity for in-plane graphite. The universal graphene conductivity $\sigma_0$ 
is also shown. The insert displays the infrared regime. b)The Casimir force between graphenes as a function of distance with constant and dynamical conductivities. 
The force is normalized to the one for perfect conductors.}
\end{figure}

Direct measurements of the graphene conductivity have been done at 
photonic energies below 3 eV so far\cite{Nair:2008}. Experiments appropriate for higher regimes have not been currently reported. At the same time, investigating $\sigma$ 
theoretically is one of the issues at the forefront of graphene science. Recent {\it ab initio} calculations indicate that the in-plane optical properties of graphite and 
graphene are very similar over a wide range of frequencies~\cite{Guo:2004,Marinopolous:2004}. Experimentally it was also reported that the optical conductivity of graphite per 
graphene sheet is very close to the universal $\sigma_0$ value of an isolated graphene~\cite{Kuzmenko:2008} for low optical frequencies. Thus a viable approach for further 
investigating the Casimir force between graphenes is to use the in-plane optical data of graphite and transpose it to graphene. 

Results from {\it ab initio} calculations for graphite have been mapped to a series of Lorentz oscillators with 
a Drude term, whose parameters fit previous graphite measurements\cite{Djurisic:1999} between $0.1\ eV$ and $40\ eV$. A plot of the in-plane conductivity as a function of photon 
energies is 
shown in Fig.(\ref{condgraph}). We note that in the infrared spectrum $\sigma$ of the two systems are different - insert of Fig.(\ref{condgraph}). 
For graphite, the conductivity exhibits a Drude-like behavior from 
intraband transitions, while the conductivity for graphene stays constant. This has also been observed experimentally~\cite{Kuzmenko:2008} and it can be explained in terms of 
the electronic structures of the two systems. The onset of the Drude-like term in graphite is related to the splitting of the 
energy bands and their becoming slightly parabolic due to the interlayer interaction, while the constant $\sigma_0$ in graphene originates from cancelations occurring between 
the intraband and interband transitions due to the linear in $k$ energy bands as shown in the Appendix. Thus for the calculations here, we modify the fitted
 model\cite{Djurisic:1999}  
for photon energies in the infrared region (below $0.05\ eV$) by requiring $\sigma=\sigma_0$ as displayed in the insert of Fig.(\ref{condgraph}).

For larger photonic energies, the in-plane graphite conductivity is mainly determined by the single graphene properties. 
The Lorentz oscillator model shows that $Re(\sigma)$ stays relatively constant in the low optical regime (up to $3 \ eV$), which means that the universal graphene conductivity has not been affected significantly 
by the presence of the other graphene layers. This is in agreement with previous experimental findings\cite{Kuzmenko:2008}. 
Also, the two peaks in $Re(\sigma)$ that appear around $5\ eV$ and $15\ eV$ are related to $\pi-\pi^*$ and $\sigma-\sigma^*$ electron transitions 
for an isolated graphene~\cite{Guo:2004,Marinopolous:2004}, respectively. A sizable imaginary part of the conductivity also appears after $3\ eV$. 

Using this model, the force between two parallel 
graphenes is found via Eq.(\ref{SPS4}). We show a plot of the normalized to perfect conductors force per unit area as a function of distance in Fig.(\ref{force2plates}) with 
$\sigma$ for graphite with the modification of $\sigma=\sigma_0$ for energies 
less than $0.05 \ eV$ and for $\sigma=\sigma_0$ over the entire range. The plot shows that at longer 
distances the force approaches the one given with a constant conductivity, but at shorter distances higher photon frequency modes contribute to an increasingly larger force.


\section{Conclusion\label{Conclusion}}

We have studied the Casimir force between parallel infinitely thin sheets in free space. The particular absorption optical properties are taken into account via the 
conductivities of each sheet. 
The derived expressions rely on generalized Fresnel reflection coefficients obtained with an iterative procedure using the Dyadic Green's function. This is especially convenient 
since it is applicable to a system of $N$ sheets. The theory is applied to the case of graphene/graphene Casimir interaction in order to 
study how this fundamental effect depends on the graphene optical response and the distance separation. The graphene conductivity is described with a model based on the 
low energy band structure first. In this case, we find that the Casimir 
force obeys the same distance dependence as the force between two perfect conductors, but it is much smaller in magnitude due to the graphene transparency. These results are 
directly related to the existence of a constant graphene conductivity $\sigma_0=e^2/(4\hbar)$ over the optical range of photon energies. This universal value translates into a 
mutual Casimir force that depends only on the electron charge and the distance. Because of its transparency, the interaction between two graphenes is not affected 
significantly when more graphenes are present.  

The graphene conductivity is also calculated using a 
model based on {\it ab initio} calculations and appropriate for in-plane graphite optical data due similarities between the two systems. 
For graphite, $\sigma$ however, has to be modified in the infrared 
photon energy region in order to reflect experimental and theoretical results for the existence of a constant graphene conductivity. This is important for the interaction in the 
limit of large separations.

Finally, we comment that the graphene conductivity might be influenced by other factors such as electron-phonon interaction, electron correlation effects and the presence of 
exciton. Efforts to understand and quantify these effects theoretically have just begun~\cite{Trevisanutto:2010}. Experimental research is also needed to validate such studies. 
Thus it would be interesting to explore in the future other models for the conductivity of graphene in relation to their mutual Casimir interaction.

\section{Acknowledgements}

We acknowledge financial support from the Department of Energy under contract DE-FG02-06ER46297.

\appendix
\section{Conductivity Model}

\begin{figure}[ht]
\begin{centering}
\includegraphics[scale=0.35]{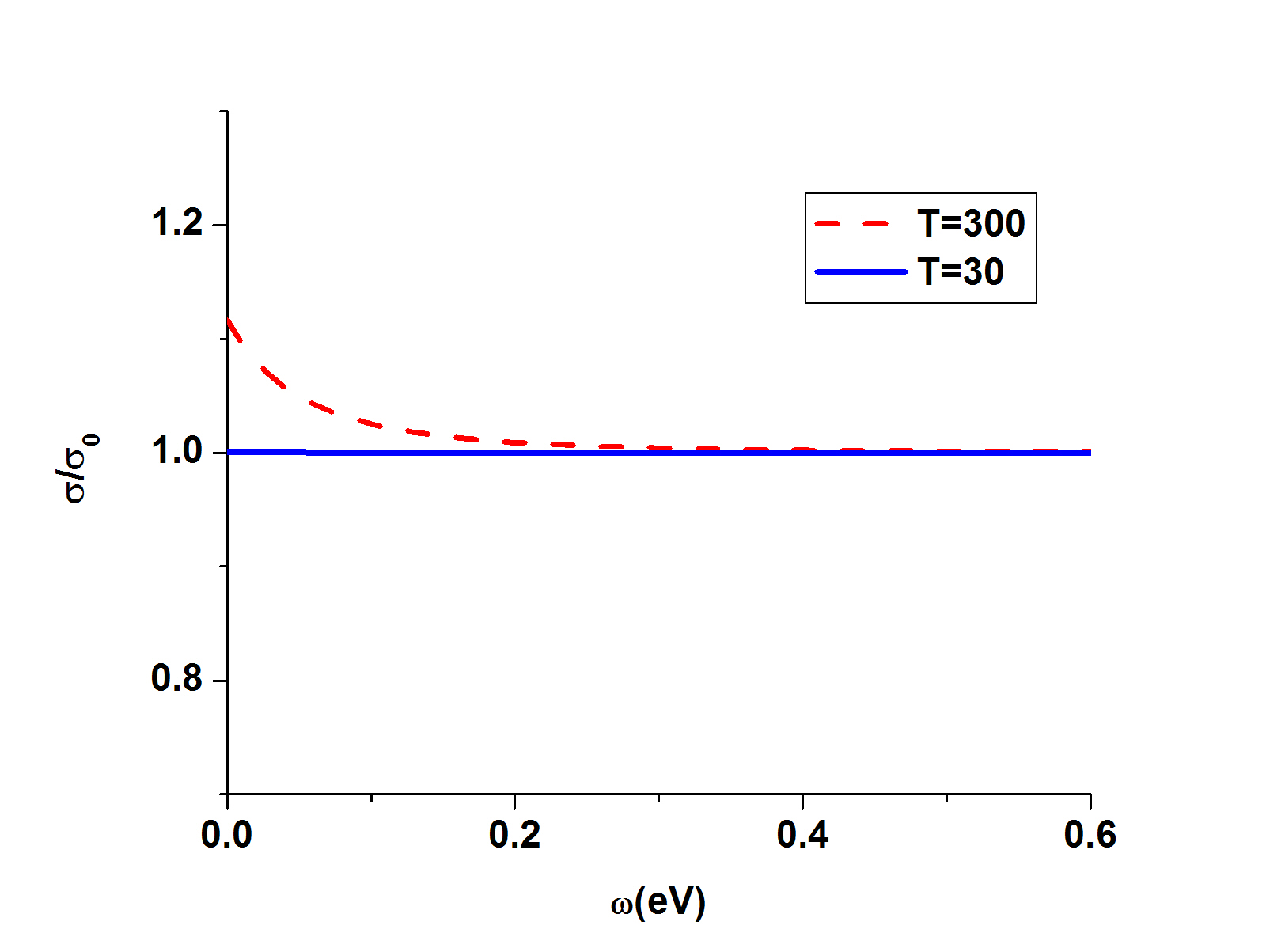}
\caption{(Color online)  Graphene conductivity $\sigma(i\omega)$ in units of $\sigma_0=e^2/(4\hbar)$ vs. frequency is given at two temperatures, $T=300\ K$ and $T=30\ K$.  The scattering rate is
$\Gamma=0.1\ eV$.}
\label{conductivity}
\end{centering}
\end{figure}
The conductivity of graphene can be modeled using the low energy electron excitations which obey a linear momentum energy dispersion relation $\epsilon=\pm v_{F}k$, 
where $v_F\approx c/300$, and $k$ is the magnitude of the two-dimensional wave vector\cite{Bordag:2009}. Within the Kubo formalism, the conductivity is expressed\cite{Hanson1:2008,Falkovsky:2007} using 
\begin{eqnarray}
\sigma(\omega,\Gamma)=\sigma_{intra}(\omega,\Gamma)+\sigma_{inter}(\omega,\Gamma),\nonumber \\
\sigma_{intra}(\omega,\Gamma)=-\frac{ie^2}{\pi\hbar^2(\omega+i\Gamma)}\int_0^\infty\epsilon d\epsilon\left(\frac{\partial f(\epsilon)}{\partial\epsilon}-
\frac{\partial f(-\epsilon)}{\partial\epsilon}\right),\nonumber \\
\sigma_{inter}(\omega,\Gamma)= \frac{ie^2(\omega+i\Gamma)}{\pi\hbar^2} \int_0^\infty d\epsilon\frac{f(-\epsilon)-f(\epsilon)}{(\omega+i\Gamma)^2-4(\epsilon/\hbar)^2} ,
\label{A1}
\end{eqnarray}
where $f(\epsilon)=1/[\exp(\epsilon/k_BT)+1)]$ is the Fermi-Dirac distribution function and $\Gamma$ is a damping parameter which accounts for physical processes contributing to 
broadening of the optical spectrum. $\sigma_{intra}(\omega,\Gamma)$ is the intraband contribution to the conductivity which is found to be
\begin{equation}
\sigma_{intra}(\omega,\Gamma)=\frac{i2e^2k_BT\ln(2)}{\pi\hbar(\omega+i\Gamma)},
\label{A2}
\end{equation}
while the intraband contribution to the conductivity is calculated as
\begin{equation}
\sigma_{intra}(\omega,\Gamma)=-\frac{ie^2(\omega+i\Gamma)}{8\pi k_BT}\int_0^\infty dx\frac{\tanh(x)}{x^2-\left(\frac{\hbar(\omega+i\Gamma)}{4k_BT}\right)^2}.
\label{A3}
\end{equation}
In the small temperature limit, A2 and A3 result in $\sigma$ being a constant - $\sigma_0=e^2/(4\hbar)$. Our sample plot of $\sigma(i\omega,\Gamma)/\sigma_0$ vs. $\omega$, however, 
shows that even at higher temperatures, the conductivity 
does not differ much from the universal constant value. Our calculations also show that the temperature entering through the conductivity has little effect on the Casimir force. 
Indeed, the force at 300 K differs by less than $1\%$ as compared to the one for 0K.


\begin{thebibliography}{99}

\bibitem{Casimir:1948}
H.B.G. Casimir, Proc. K. Ned. Akad. Wet. {\bf 51}, 793 (1948).

\bibitem{Mostepanenko:2009}
V.M. Mostepanenko, M. Bordag, G.L. Klimchitskaya and U. Mohideen, {\it Advances in the Casimir Effect} (Oxford University Press, Oxford (2009)).

\bibitem{Chan1:2001}
H.B. Chan, V.A. Aksyuk, R.N. Kleiman, D.J. Bishop, and F. Capasso,  Phys. Rev. Lett. {\bf 87}, 211801 (2001).

\bibitem{Chan:2001}
H.B. Chan, et. al.,  Science {\bf 291}, 1941 (2001).

\bibitem{Lamoreaux:1997}
S.K. Lamoreaux, Phys. Rev. Lett. {\bf 78}, 5 (1997).

\bibitem{Mohideen:1998}
U. Mohideen and A. Roy, Phys. Rev. Lett. {\bf 81}, 4549 (1998). 

\bibitem{Dresselhaus_book} 
R. Saito, G. Dresselhaus, and M.S. Dresselhaus, {\it Physical Properties of Carbon Nanotubes} (Imperial College Press, London, 1998).

\bibitem{Neto:2009} 
A.H. Castro Neto, F. Guinea, N.M.R. Peres, K.S. Novoselov, and A.K. Geim, Rev. Mod. Phys. {\bf 81}, 109 (2009).

\bibitem{Novoselov:2004}
K.S. Novoselov, A.K. Geim, S.V. Morozov, D. Jiang, Y. Zhang, S.V. Dubonos, I.V. Grigorieva, and A.A. Firsov,  Science {\bf 360}, 666 (2004). 

\bibitem{Novoselov:2005}
K.S. Novoselov, D. Jiang, F. Schedin, T.J. Booth, V.V. Khotkevich, S.V. Morozov, and A.K. Geim,  Proc. Nat. Acd. Sci. {\bf 102},
10451 (2005).

\bibitem{Geim:2007}
A.K. Geim, and K.S. Novoselov, Nat. Mat. {\bf 6}, 183 (2007).

\bibitem{Lin:2010}
Y.-M. Lin, C. Dimitrakopoulos, K.A. Jenkins, D.B. Farmer, and H.-Y. Chiu, Science {\bf 327}, 662 (2010).

\bibitem{Chen:2009}
C. Chen et al, Nature Nanotech. {\bf 4}, 861 (2009).

\bibitem{Klimchitskaya:2009}
G.L. Klimchitskaya, U. Mohideen, and V.M. Mostepanenko, Rev. Mod. Phys. {\bf 81}, 1827 (2009).

\bibitem{Bordag:2006}
M. Bordag, B. Geyer, G.L. Klimchitskaya, and V.M. Mostepanenko, Phys. Rev. B {\bf 74}, 205431 (2006). 

\bibitem{Bordag:2009}
M. Bordag, I. V. Fialkovsky, D.M. Gitman, and D.V. Vassilevich, Phys. Rev. B {\bf 80}, 245406 (2009).

\bibitem{Landau:1980}
L.D. Landau and E.M. Lifshitz,   {\it Statistical Physics, Ch. XII}  (Butterworth Heinemann, Oxford (1980)).

\bibitem{Kubo:1957}
R. Kubo, J. Phys. Soc. Jap. {\bf 12}, 570 (1957).

\bibitem{Knoll:2001}
L. Kn\"oll, S. Scheel, and D.-G. Welsch, {\it QED in Dispersing and Absorbing Dielectric Media} in ``Coherence and Statistics of Photons and Atoms,''
(Jan Perina, John Wiley \& Sons, Inc, New York (2001)).

\bibitem{Tomas:2002}
M.S. Tomas, Phys. Rev. A {\bf 66}, 052103 (2002).

\bibitem{Raabe:2003}
C. Raabe, L. Kn\"oll, and D.-G. Welsch, Phys. Rev. A {\bf 68}, 033810 (2003).

\bibitem{Hanson:2008}
G.W. Hanson, IEEE Trans. Anten. Prop. {\bf 56}, 747 (2008).

\bibitem{Tai:1993}
Chen-To Tai, {\it Dyadic Green Functions in Electromagnetic Theory, Ch. 11} (IEEE Press, Piscataway (1993)).

\bibitem{Cheng:1986}
D.H.S. Cheng, Electromagnetics {\bf 6}, 171 (1986).

\bibitem{Abrikosov:1963}
A.A. Abrikosov, L.P. Gorkov, and I.E. Dzyaloshinski, {\it Methods of Quantum Field Theory in Statistical Physics, Ch. 6} (Dover Publications, Inc, New York (1963)).
 
\bibitem{Ellingsen:2007}
S.A. Ellingsen,  J. Phys. A:Math Theor. {\bf 40}, 1951 (2007). 


\bibitem{Santos:2009}
G. Gomez-Santos, Phys. Rev. B {\bf 80}, 245424 (2009).

\bibitem{Gusynin:2006}
V.P. Gusynin, S.G. Sharapov, and J.P. Carbotte,  J.Phys.: Condens. Matter {\bf 19}, 026222 (2007).

\bibitem{Falkovsky:2007}
L.A. Falkovsky and A.A. Varlamov,  The Eur. Phys. J. B {\bf 56}, 281 (2007). 

\bibitem{Nair:2008}
R.R. Nair, P. Blake, A.N. Grigorenko, K.S. Novoselov, T.J. Booth, T. Stauber, N.M.R. Peres, and A.K. Geim, Science {\bf 320}, 1308 (2008).

\bibitem{Abramowitz:1964}
M. Abramowitz and I.A. Stegun, {\it Handbook of Mathematical Functions, p. 263} (Dover Publicatons Inc., New York (1964)).

\bibitem{Guo:2004}
G.Y. Guo, K.C. Chu, D. S. Wang, and C. G. Duan, Phys. Rev. B {\bf 69}, 205416 (2004).

\bibitem{Marinopolous:2004}
A.G. Marinopoulos, L. Reining, A. Rubio, and V. Olevano, Phys. Rev. B {\bf 69}, 245419 (2004).

\bibitem{Kuzmenko:2008}
A.B. Kuzmenko, E.van Heumen, F. Carbone, and D. van der Marel, Phys. Rev. Lett. {\bf 100}, 117401 (2008).

\bibitem{Djurisic:1999}
A.B. Djuri\v{s}i\'{c} and E.H. Li,  J. of Appl. Phys. {\bf 85}, 7404 (1999).

\bibitem{Hanson1:2008}
G.W. Hanson, J. of Appl. Phys. {\bf 103}, 064302 (2008).

\bibitem{Trevisanutto:2010}
P.E. Trevisanutto, M. Holzmann, M. Cote, and V. Olevano, Phys. Rev. B {\bf 81}, 121405(R) (2010).

\end{thebibliography}
\end{document}